\documentclass[onecolumn,aps,pra,showpacs]{revtex4-1}

\usepackage{bm}

\usepackage{graphicx}
\usepackage{epstopdf}

\begin{document}
\draft
\preprint{}

\title{A new approach for strong-field ionization}

\author{V.D. Rodr\'iguez}
\affiliation{Departamento de F\'{\i}sica, FCEyN, UBA, Ciudad Universitaria Pabell\'on 1, 1428 Buenos Aires, Argentina}
\affiliation{IFIBA CONICET, FCEyN, UBA, Ciudad Universitaria Pabell\'on 1, 1428 Buenos Aires, Argentina}
\email[E-mail me at: ]{vladimir@df.uba.ar}
\begin{abstract}
A model describing the electronic transitions in  an atom subject to a strong high frequency laser pulse  is proposed in the velocity gauge. The model accounts for the initial state coupling with the the remaining discrete  and continuum spectra. Continuum-continuum transitions are also taken into account following well known strong-field approximation. A single integro-differential equation for the initial state amplitude has been obtained. Exact numerical solutions of this equation are  computed and compared with the full time dependent Schr{\"o}dinger equation solution for Hydrogen atoms. The initial state amplitude is well described by two versions of the model: with and without the continuum-continuum coupling model. However the electron energy spectrum is only well described when the model accounts for the  coupling between continuum states. Improvement on SFA arise from using  Coulomb final wave function as well as regarding the initial state probability depletion. 
\end{abstract}

\pacs{42.50 Hz , 32.80.Rm , 32.80.Fb }

\maketitle

\section{Introduction}

Most of the theoretical work on atomic ionization by an intense and short laser pulse can be related to the Strong Field Approximation (SFA) approach  \cite{keldysh65,faisal73,reiss80} (see also \cite{milonni89}). According to this theory, after ionization takes place, the emitted electron evolves under the interaction with the laser field radiation. This theory has proved to be appropriate to deal with the detachment problem, in which the ionized electron is emitted by a negative ion which, after ionization takes place, becomes a neutral atom \cite{gribakin}. In this case the photoelectron once ionized evolves only under the interaction with the laser field. However, the theory has been formulated under the assumption that no appreciable depletion of the system initial state takes place owing to the laser field \cite{lewenstein}.

For neutral atom ionization, a closely related theory has been formulated to account for the  long range Coulomb interaction in the final state, namely, the so called Coulomb-Volkov approximation (CV2$^{-}$) \cite{duchateau01,duchateau02}. According to this theory, after ionization has taken place, the emitted electron evolves under both the interactions with the residual ion and the laser radiation. The spectra obtained with the CV2$^{-}$ approximation has been obtained for the case in which the laser frequency is greater than the ionization potential and for not too strong laser fields amplitudes.

The CV2$^{-}$ approach provides an accurate description of the ionization spectra and, in particular, it accounts for the ATI peaks of the spectrum when the mentioned conditions hold and the laser frequency is larger than the ionization potential. When the laser frequency is lower than the ionization potential, a modified version of CV2$^{-}$ \cite{rodriguez04} provides the correct description of the physical ionization process, but the theory still fails under strongly non perturbative conditions.

Several works that treat the strong laser field in a non-perturbative way, have been published \cite{gayet05,babiarz01}.

In order to abandon the perturbative conditions, the different theories have based their models on the inclusion of the depletion of the initial atomic state that occurs when the stronger laser fields are taken into consideration.

The depletion of the initial state allows for the quick transfer to the continuum of the population of the initial state and the posterior rescattering process that characterizes the ionization regimes at higher laser field intensities.

For instance, the results obtained with the renormalized CV2$^{-}$ theory \cite{gayet05}, have been able to correct the magnitude of the spectrum, {\it i.e.}, the magnitude of the principal and secondary ATI peaks and the background computed from the CV2$^{-}$ theory. The spectra computed with the original CV2$^{-}$ theory has shown to be considerably improved after the renormalization process.  Unfortunately, the theory neglects the initial state coupling with other discrete states, and  a Weiskopf-Wigner approximation was employed \cite{gayet05}.

On the other hand, the theory proposed in \cite{babiarz01}, accounts for the depletion of the initial atomic state by including the coupling between the initial state and the continuum states of the atom. However the authors oversimplify the problem in order to make the calculation as analytical as possible. Thus, a simple model for the coupling between the initial state with the continuum and discrete part of the spectrum was proposed, though the continuum-continuum coupling is considered in the model. Again, no other discrete states are considered.

The present alternative approach uses real coupling matrix elements. We will show the model to be unitary. The model is computed in the velocity gauge, and the probability amplitudes of the states involved are obtained from the exact solution of the resulting integro-differential equation (IDE) after the main approximations of the model are applied to the initial system of differential equations in the transitions amplitudes. We have tested the model for non-perturbative situations where the strength of the field has been raised to levels of the order of the atomic unity.

In this paper we focus on the case in which the photon energy is greater than the ionization potential, and leave the case in which the photon energy is lower than the ionization potential, where the resonance phenomenon has to be taken into account, to be considered in further work.

In section II we present the theory behind the present unitary model. In section  III we present the results obtained with the model for the case where the photon energy is greater than the ionization potential and compare them with exact time dependent Schr\"odinger equation  (TDSE) simulations. Sec. IV presents some of our conclusions.

Atomic units are used otherwise stated.

\section{Theory}

\subsection{The model}

In the one-active electron approximation, the laser-atom Hamiltonian computed in the velocity gauge, is given by
\begin{equation}
\qquad\qquad\qquad\qquad \hat{H}=\hat{H}_{0} + \hat{V}
\label{TotalHamiltonian}
\end{equation}
where
\begin{equation}
\qquad\qquad\qquad\qquad \hat{H}_{0}=\frac{\mathbf{\hat{p}}^{2}}{2}+\hat{V}_{T}(\mathbf{r})
\label{AtomicHamiltonian2}
\end{equation}
is the non relativistic atomic Hamiltonian in the absence of the laser field, $\hat{V}_{T}(\mathbf{r})$ represents the target atomic  potential, and
\begin{equation}
\qquad\qquad\qquad\qquad \hat{V}=\mathbf{A} \cdot \mathbf{\hat{p}}+\frac{1}{2}\mathbf{A}^{2}
\label{InteractionHamiltonian}
\end{equation}
is the laser-atom interaction Hamiltonian. Here $\mathbf{\hat{p}}$ is the electron momentum operator and $\mathbf{A}$ the laser pulse vector potential field, which under the dipolar approximation may be considered only time dependent, {\it i.e.}, $\mathbf{A}(\mathbf{r},t)\approx \mathbf{A}(t)$. As the $\frac{1}{2}\mathbf{A}(t)^{2}$ term is only time dependent, it can be exactly accounted for through a time-dependent phase factor \cite{bauer06}, therefore, we dismiss this term from now on.\\
The TDSE for the ket state $| \psi (t)\rangle$ of the atomic expansion is then given by
\begin{equation}
\qquad\qquad\qquad \qquad i\frac{\partial{|\psi (t)}\rangle}{\partial{t}} =H(t)|\psi (t)\rangle.
\label{SEq}
\end{equation}
With an initial condition like
\begin{equation}
\qquad\qquad\qquad\qquad |\psi (0)\rangle =
|\phi (0)\rangle = |i\rangle,
\label{SEq}
\end{equation}
the TDSE represents an initial value problem. The solution of the TDSE can be obtained either by a numerical procedure or by analytical approximations.
We perform the expansion of the time-dependent state ket in eigenkets of the atomic Hamiltonian and separate the initial state	 $|i\rangle$ with energy $\varepsilon_{i}$, obtaining thus
\begin{equation}
|\psi(t)\rangle = a_{i}(t)e^{-i\varepsilon_{i}t} |i\rangle + \sum_{n\neq i} a_{n}(t)e^{-i\varepsilon_{n}t} |n\rangle + \int d\mathbf{k}																				 a_{\mathbf{k}}(t)e^{-i\varepsilon_{k}t}|\mathbf{k}\rangle,
\label{StateKetExpansion}
\end{equation}
where $|n\rangle$ $(n \neq i)$ and $|\mathbf{k}\rangle$ are the eigenkets corresponding to the discrete  $\varepsilon_{n}$ and continuous $\varepsilon_{k}$ energy levels, respectively. The continuum eigenkets are assumed to be normalized in the momentum scale.	 
When the last expansion is replaced into the TDSE and projected onto each eigenket considered in the expansion, the following  system of coupled first order differential equations for the expansion amplitudes is obtained:
\begin{eqnarray}
\qquad\qquad \qquad \dot{a}_{i}(t)=-i \sum_{n\neq i} V_{in}(t) e^{-i(\varepsilon_{n}-\varepsilon_{i})t}a_{n}(t)-i \int d\mathbf{k}																				 V_{i\mathbf{k}(t)} e^{-i(\varepsilon_{k}-\varepsilon_{i})t}a_{\mathbf{k}}(t)\label{DiAmplitude}
\end{eqnarray}
\begin{eqnarray}
\qquad \qquad \qquad \dot{a}_{n}(t)=-i V_{ni}(t) e^{-i(\varepsilon_{i}-\varepsilon_{n})t}a_{i}(t)-i \sum_{m\neq i} V_{nm}(t) e^{-i(\varepsilon_{m}-\varepsilon_{n})t}a_{m}(t)\nonumber\\
\qquad \qquad -i \int d\mathbf{k}V_{n \mathbf{k}}(t) e^{-i(\varepsilon_{k}-\varepsilon_{n})t}a_{\mathbf{k}}(t)\label{DniAmplitude}
\end{eqnarray}
\begin{eqnarray}
\qquad \qquad \qquad \dot{a}_{\mathbf{k}}(t)=-i V_{\mathbf{k}i}(t) e^{-i(\varepsilon_{i}-\varepsilon_{k})t}a_{i}(t)-i \sum_{n\neq i} V_{\mathbf{k}n}(t) e^{-i(\varepsilon_{n}-\varepsilon_{k})t}a_{n}(t)\nonumber\\
\qquad \qquad -i \int d\mathbf{k^{\prime}}V_{\mathbf{k}\mathbf{k}^{\prime}}(t)
e^{-i(\varepsilon_{k^{\prime}}-\varepsilon_{k})t}a_{\mathbf{k}^{\prime}}(t)
\label{DKiAmplitude}
\end{eqnarray}
Two main approximations are considered in this model. First
the only  couplings accounted for are those between the initial state and the rest of the atomic spectrum and between continuum states.
Second, we approximate the continuum-continuum coupling by the expression
\begin{equation}
\qquad\qquad \qquad V_{\mathbf{k}\mathbf{k}^{\prime}}(t)\approx\mathbf{A}(t)\cdot \mathbf{k}\delta(\mathbf{k}-\mathbf{k}^{\prime}).\label{CC-coupling1}
\end{equation}
Therefore, the modified (\ref{DniAmplitude}) is given by
\begin{equation}
\qquad\qquad \dot{a}_{n}(t)=-i \mathbf{\hat{p}}_{\mathrm{ni}}\cdot\mathbf{A}(t) e^{-i(\varepsilon_{i}-\varepsilon_{n})t}a_{i}(t)\label{DnAmplitude}.
\end{equation}
This equation can be formally integrated to obtain
\begin{equation}
\qquad\qquad a_{n}(t)=-i\int_0^{t}dt^{\prime}\mathbf{\hat{p}}_{\mathrm{ni}}\cdot
\mathbf{A}(t^{\prime}) e^{-i(\varepsilon_{i}-\varepsilon_{n})t^{\prime}}a_{i}(t^{\prime}).
\label{nAmplitude1}
\end{equation}
Using (\ref{CC-coupling1}), the modified version of (\ref{DKiAmplitude}) is
\begin{eqnarray}
\qquad
\dot{a}_{\mathbf{k}}(t)= - i \mathbf{\hat{p}}_{\mathrm{\mathbf{k}i}}
\cdot\mathbf{A}(t) e^{-i(\varepsilon_{i}-\varepsilon_{k})t}
a_{i}(t)
 - i \sum_{n\ne i}\mathbf{\hat{p}}_{\mathrm{\mathbf{k}n}}
\cdot\mathbf{A}(t) e^{-i(\varepsilon_{n}-\varepsilon_{k})t}
a_{n}(t)\nonumber\\
-i \int\mathrm{d}\mathbf{k}'
\mathbf{k}\cdot\mathbf{A}(t)\delta(\mathbf{k}-\mathbf{k}^{\prime})
e^{-i(\varepsilon_{k'}-\varepsilon_{k})t}a_{\mathbf{k}'}(t)\nonumber\\
=- i \mathbf{\hat{p}}_{\mathrm{\mathbf{k}i}}
\cdot\mathbf{A}(t) e^{-i(\varepsilon_{i}-\varepsilon_{k})t}
a_{i}(t)
- i \sum_{n}\mathbf{\hat{p}}_{\mathrm{\mathbf{k}n}}
\cdot\mathbf{A}(t) e^{-i(\varepsilon_{n}-\varepsilon_{k})t}
a_{n}(t) \nonumber\\
-i \mathbf{k}\cdot\mathbf{A}(t)a_{\mathbf{k}}(t),
\label{DkAmplitude}
\end{eqnarray}
this equation can be exactly integrated in terms of the initial state amplitude giving
\begin{eqnarray}
a_{\mathbf{k}}(t)=-i\int_0^{t}dt^{\prime}
\mathbf{\hat{p}}_{\mathrm{\mathbf{k}i}}(t)\cdot\mathbf{A}(t^{\prime})
e^{-i(\varepsilon_{i}-\varepsilon_{k})t^{\prime}}a_{i}(t^{\prime})
e^{-i \mathbf{k}\cdot\bm \alpha (t,t^{\prime})}\nonumber\\
\qquad -i\sum_{n\ne i}\int_0^{t}dt^{\prime}
\mathbf{\hat{p}}_{\mathrm{\mathbf{k}n}}(t)\cdot\mathbf{A}(t^{\prime})
e^{-i(\varepsilon_{n}-\varepsilon_{k})t^{\prime}}a_{n}(t^{\prime})
e^{-i \mathbf{k}\cdot\bm \alpha (t,t^{\prime})},
\label{kAmplitude1full}
\end{eqnarray}
where
\begin{equation}
\qquad\qquad\qquad \bm \alpha (t,t^{\prime})=\int_{t^{\prime}}^{t}dt^{\prime\prime}
\mathbf{A}(t^{\prime\prime}).
\label{alpha}
\end{equation}
This equation contains both contributions, direct $i\to \mathbf{k}$  and two-step $i\to n$ followed by $n \to \mathbf{k}$ transitions. In order to proceed we keep from this equation only the leading direct contribution,
\begin{eqnarray}
a_{\mathbf{k}}(t)=-i\int_0^{t}dt^{\prime}
\mathbf{\hat{p}}_{\mathrm{\mathbf{k}i}}(t)\cdot\mathbf{A}(t^{\prime})
e^{-i(\varepsilon_{i}-\varepsilon_{k})t^{\prime}}a_{i}(t^{\prime})
e^{-i \mathbf{k}\cdot\bm \alpha (t,t^{\prime})}
\label{kAmplitude1}
\end{eqnarray}

By replacing (\ref{nAmplitude1}) and (\ref{kAmplitude1})	in	 (\ref{DiAmplitude}), the following integro-differential	 equation (IDE) for the initial state amplitude is obtained:
\begin{eqnarray}
\qquad\qquad\qquad \dot{a}_{i}(t) = - \int_{0}^{t}dt^{\prime}K(t,t^{\prime})a_{i}(t^{\prime}).
\label{ai-IDE01}
\end{eqnarray}
where the kernel of the IDE is given by
\begin{eqnarray}
\qquad\qquad\qquad K(t,t^{\prime})= \sum_{n\neq i} [\mathbf{\hat{p}}_{in}\cdot\mathbf{A}(t)]
[\mathbf{\hat{p}}_{ni}\cdot\mathbf{A}(t^{\prime})]
e^{-i\varepsilon_{n}(t-t^{\prime})}+ \nonumber\\
\qquad\qquad\qquad \int[\mathbf{\hat{p}}_{i\mathbf{k}}\cdot\mathbf{A}(t)]
[\mathbf{\hat{p}}_{\mathbf{k}i}\cdot\mathbf{A}(t^{\prime})]
e^{-i\varepsilon_{k}(t-t^{\prime})}e^{-i \mathbf{k}\cdot\bm \alpha (t,t^{\prime})}
\label{kernel}
\end{eqnarray}
As an application of the model presented above, we consider the usual case, in which the laser field is linearly polarized, for example, in the $z$-direction. In this case, the mean value of the momentum operator between the initial state and the \emph{n}-discrete and $\mathbf{k}$-continuum states are respectively $p_{z_{ni}}=\left\langle n|\hat{p}_{z}|i\right\rangle$ and
 $p_{z_{\mathbf{k}i}}=\left\langle \mathbf{k}|\hat{p}_{z}|i\right\rangle$. In this case (\ref{kernel}) may be simplified to
\begin{equation}
\qquad \dot{a}_{i}(t)=(-i)^{2}A(t)e^{i\varepsilon_{i}t}\int_{0}^{t}dt^{\prime }A(t^{\prime})e^{-i\varepsilon_{i}t^{\prime}}h(t,t^{\prime})a_{i}(t^{\prime}),
\label{ai-IDE1}
\end{equation}
with the function in the kernel defined as
\begin{equation}
\qquad\qquad\qquad h(t,t^\prime)=h_{D}(t-t^{\prime})+h_{C}(t,t^{\prime})
\label{hV}
\end{equation}
In this last equation
\begin{equation}
\qquad\qquad h_{D}(t-t^{\prime})=\sum_{n\neq i}|p_{z_{ni}}|^{2}e^{-i\varepsilon_{n}(t-t^{\prime})}\\
\label{hDV}
\end{equation}
and
\begin{equation}
\qquad\qquad h_{C}(t,t^{\prime})=\int d\mathbf{k}|p_{z_{\mathbf{k}i}}|^{2}e^{-i\varepsilon_{k}(t-t^{\prime})}e^{-i k \cos\theta_{k}\alpha(t,t^{\prime})}.
\label{hCV}
\end{equation}
The subindexes D and C, in (\ref{hV}), stand for discrete and continuous contributions, respectively.
The energy levels diagrams of figure 1 display the different coupling schemes considered in this work. The scheme presented in 1a) accounts for the depletion of the initial state due to the transition to continuum states and for the initial state coupling with the remaining discrete part of the spectrum. In b), in addition to the couplings between the initial state and the rest of the spectrum considered in a), the continuum-continuum coupling is also taken also into account.
\begin{figure}
\includegraphics[width=14 cm,height=9 cm]{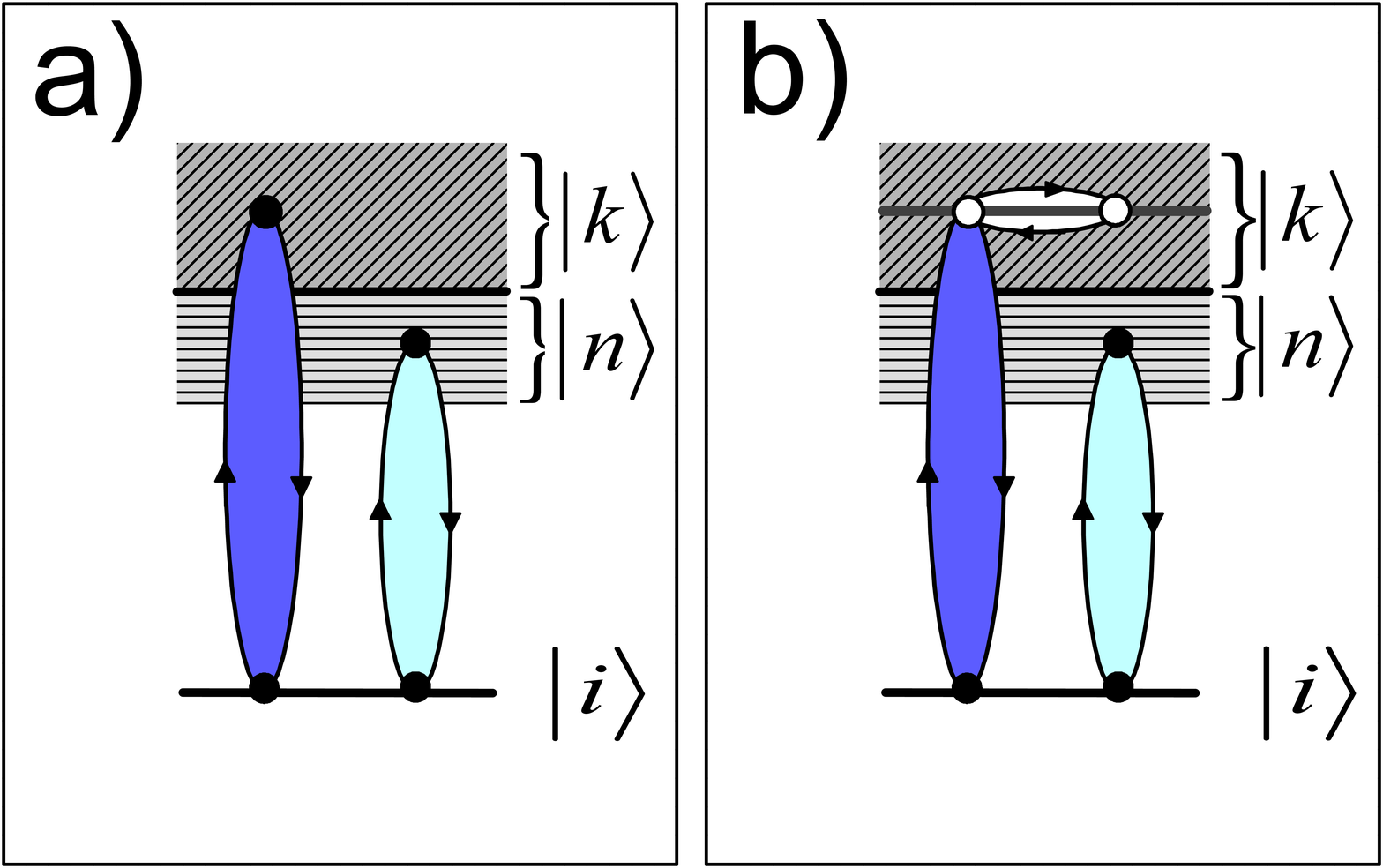}
\begin{caption}{Coupling schemes considered in this work. In a) 'vertical' transitions between the initial state and continuum or discrete states and in b) both 'vertical' and 'horizontal' continuum-continuum transitions. Color on-line.}
\end{caption}
\end{figure}
The functions $h_{C}$ and $h_{D}$ represent the coupling between the initial state and the discrete and continuum atomic spectra, respectively, computed both in the velocity gauge.
The case in which the the continuum-continuum coupling is not taken into consideration in the model, may be easily obtained from the full coupled model by setting the function $\alpha(t,t^{\prime})$ equal to zero. With a change of variables $({t,t^{\prime}})\rightarrow ({\alpha,t^{\prime \prime}=t-t^{\prime}})$, it is only necessary to compute the function $h_{C}(\alpha,tt^{\prime})$.
This function  can be computed as the integral
\begin{equation}
\qquad\qquad h_{C}(\alpha,t^{\prime \prime})=\ \int dk e^{-i\varepsilon_{k}(t^{\prime \prime})}\beta_{0}(\alpha,k).
\label{hCVIntegral}
\end{equation}

We apply this model to the Hydrogen atom in its ground state interacting with a linearly polarized laser pulse.
For this particular case, the function $\beta_{0}$ that appears in (\ref{hCVIntegral}) is analytical. For the case that takes into account the continuum-continuum coupling, it is given by
\begin{eqnarray}
\qquad  \beta_{0}(\alpha,k)=\int d\Omega_k k^{2}|p_{z_{\mathbf{k}0}}|^{2}e^{-i k \cos\theta_{k}\alpha(t,t^{\prime})}\\
=\frac{32e^{\frac{\pi-4\arctan(k)}{k}}k\sinh^{-1}(\frac{\pi}{k})}{3(1+k^{2})^{3}}\frac{2\alpha k\cos(\alpha k)+(-2+\alpha^{2}k^{2})\sin(\alpha k)}{\alpha^{3}k^{3}}.
\label{beta0alphak}
\end{eqnarray}
If the continuum-continuum coupling is not taken into account, the parameter $\alpha(t,t^{\prime})$ must be set to zero and the corresponding $\beta_{0}$ function becomes then
\begin{eqnarray}
\qquad\qquad \qquad \qquad \qquad \beta_{0}(0,k)=\frac{32e^{\frac{\pi-4\arctan(k)}{k}}k\sinh^{-1}(\frac{\pi}{k})}{3(1+k^{2})^{3}}.
\label{beta0k}
\end{eqnarray}
For deriving these equations we have used the the dipolar operator matrix element obtained with the Nordsieck method \cite{nordsieck54}:
\begin{equation}
p_{z_{\mathbf{k}0}}= \frac{4 \sqrt {2} \exp[\frac{\pi}{2k}-\frac{2\arctan k}{k}](i+\frac{1}{k})k\cos[\vartheta_k]\Gamma[1-\frac{i}{k}]}{(1+k^{2})^{3}\pi}
\label{Pzk0}.
\end{equation}
The integro-differential equation for the amplitude of the initial state, given by  (\ref{ai-IDE1}) together with the initial condition $a_{0}(0)=1$ is an IDE of the Volterra type. Its solution  has been obtained numerically using an algorithm proposed in \cite{goldfine77}, based on the Taylor expansion. A Filon's algorithm for dealing with highly oscillating functions integrals \cite{milovanovic77}, has been adapted
for the computation of the different terms of the IDE.\\

\subsection{Ionization rate}
The total ionization probability can be expressed from the square modulus of transition	 amplitude to the continuum	 with momentum $\mathbf{k}$, by integrating over the momentum space,
\begin{equation}
\qquad\qquad\qquad\qquad P_{ioni}(t)=\int d\mathbf{k}|a_{\mathbf{k}}(t)|^{2}.
\label{Pioni}
\end{equation}
The \emph{ionization rate} is then given by the formula
\begin{equation}
\qquad\qquad\qquad \dot{P}_{ioni}(t)=\int d\mathbf{k}[a_{\mathbf{k}}(t)\dot{a}_{\mathbf{k}}^*(t)+\texttt{c.c}]
\label{DPioni}
\end{equation}
where $\texttt{c.c}$ denotes the complex conjugate of the first term on the right side.
Replacing the expression for  $a_{\mathbf{k}}(t)$ given by (\ref{kAmplitude1}), after performing the corresponding temporal derivatives, we obtain (that)
\begin{eqnarray}
\dot{P}_{ioni}(t)=\int_{0}^{t}dt^{\prime }A(t)A(t^{\prime})\exp[i\varepsilon_{i}(t-t^{\prime})]a_{i}^*(t)a_{i}(t^{\prime}) \int d\mathbf{k}|p_{z_{\mathbf{k}i}}|^{2}\exp[-i\varepsilon_{k}(t-t^{\prime})] \nonumber \\
\qquad\qquad + \texttt{c.c.}
\end{eqnarray}
\begin{eqnarray}
\dot{P}_{ioni}(t)=\int d\mathbf{k} p_{z_{\mathbf{k}i}}^* A(t)\exp[i(\varepsilon_{i}-\varepsilon_{k})t] a_{0}^*(t)\nonumber
\times\int_{0}^{t}dt^{\prime }p_{z_{\mathbf{k}i}}A(t^{\prime})\exp[i(\varepsilon_{i}-\varepsilon_{k})t^{\prime}]a_{0}(t))\nonumber\\
\qquad\qquad +\texttt{c.c.}\nonumber\\
\qquad =\int_{0}^{t}dt^{\prime }A(t)A(t^{\prime})\exp[i\varepsilon_{i}(t-t^{\prime})]a_{0}^*(t)a_{0}(t^{\prime})
\int d\mathbf{k}|p_{z_{\mathbf{k}i}}|^{2}\exp[-i\varepsilon_{k}(t-t^{\prime})]+\texttt{c.c.}
\end{eqnarray}\\
Using (\ref{hCV}) the ionization rate may be expressed as
\begin{equation}
\qquad\qquad \dot{P}_{ioni}(t)=A(t)\exp(i\varepsilon_{i}t)a_{i}^*(t)\int_{0}^{t}dt^{\prime }A(t^{\prime})\exp[-i\varepsilon_{i}t^{\prime}]a_{i}(t^{\prime})h_{C}^{V}(t,t^{\prime})+\texttt{c.c.}
\label{DPioni1}
\end{equation}
The proof of the unitary property of the model based on the ionization rate given in (\ref{DPioni1}) is given in appendix A.
The numerical integration of this differential equation with initial condition $P_{ioni}(0)=0$ provides the time-dependent total ionization probability.

\section{Results}

Two versions of the present unitary model are considered in this work. In the first one denoted by $UM^{DC}$, the atomic initial bound state is coupled to both the discrete and the continuous spectrum. In the second one, here denoted by $UM^{DC+CC}$, the coupling between continuum states is incorporated to the previous model. These coupling schemes between the initial states and the rest of the atomic spectrum, are shown in figure 1 as previously mentioned.
In this work we present the results obtained for atomic Hydrogen under an intense XUV laser pulse. The laser frequency is taken to be larger than the ionization potential.
The electric field pulse is modeled by

\begin{equation}
\vec{F}(t)=
\left\{
\begin{array}{cc}
\vec{F}_{0}\sin(\omega t+\varphi) \sin^{2}(\frac {\pi t}{\tau}) \qquad & \texttt{if} \qquad t\in{(0,\tau)}\\
0 \qquad & \texttt{elsewhere}.
\end{array}
\right.
\label{pulse}
\end{equation}

The laser pulse is considered to be linearly polarized along the $z$-direction, {\it i.e.}, $\vec{F}_{0}=F_{0}\widehat{z}$.
The parameter $\tau$ stands for the duration of the pulse, the sine-square factor defines the pulse envelope, $\omega$ is the laser angular frequency and $\varphi$ the phase of the carrier.
The potential vector field amplitude $A(t)$, is computed by direct integration of (\ref{pulse}).
In what follows the field frequency of the laser pulse is taken to be $\omega=0.6$ a.u. and the pulse length is fixed in $20$ cycles.

\begin{figure}
\includegraphics[width=14 cm,height=9 cm]{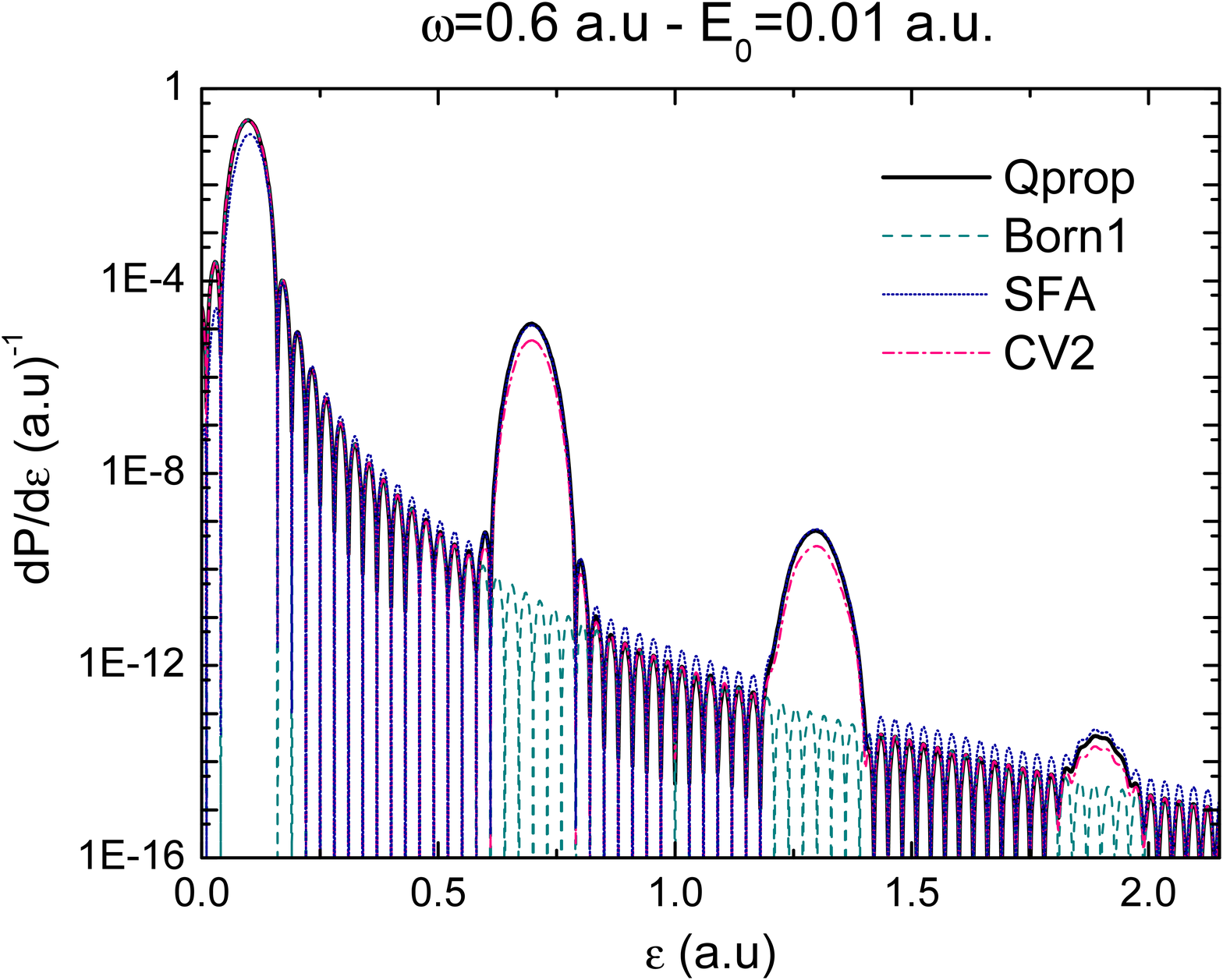}
\begin{caption}{Spectra computed under perturbative conditions: First Born, Strong Field and Coulomb-Volkov approximations computed for a laser pulse with frequency $\omega=0.6$ a.u. ($16.33 eV$), $20$ cycles of duration and amplitude equal to $E_{0}=0.01$ a.u. ($3.52\times 10^{12} W/cm^{2}$). Full lines, TDSE obtained with the Qprop code [2], dashed line, $Born 1$, dotted lines $SFA$, and dash-dotted lines CV2$^{-}$.}
\end{caption}
\label{PertB1SFACV}
\end{figure}

In figure 2 we present spectra calculations with  three  widely used theories for the study of  laser-atom interaction systems. The theories are the first Born, the SFA \cite{reiss80} and the CV2$^{-}$ \cite{duchateau02} approximations. Also, the exact numerical solution to the TDSE provided by the Qprop code \cite{bauer06} is plotted. The laser pulse used to make the calculations, has a frequency  $w=0.6$ a.u. ($16.33$ eV), 20 cycles  duration, and the field amplitude is $E_{0}=0.01$ a.u. ($3.52\times 10^{12}$  W/cm$^{2}$), characterizing a perturbative regime.

In this regime, the three theories considered provide a good description of the ionization process. In particular, the magnitude and position of the first peak of the spectrum are correctly described by the first Born and the Coulomb-Volkov approximations. The difference between the two theories is that the first Born approximation is able only to reproduce the first ATI peak and the background of the spectrum while the CV2$^{-}$ approximation, even though it underestimates the magnitude of the remaining peaks of the spectrum, is able to reproduce correctly its position, and the background. The SFA, instead, reproduces correctly the position of the first ATI peak, though slightly underestimates its height. The magnitude and position of the remaining peaks and the background of the spectrum are, for this particular case, estimated  correctly by the SFA.

\begin{figure}
\includegraphics[width=14 cm,height=9 cm]{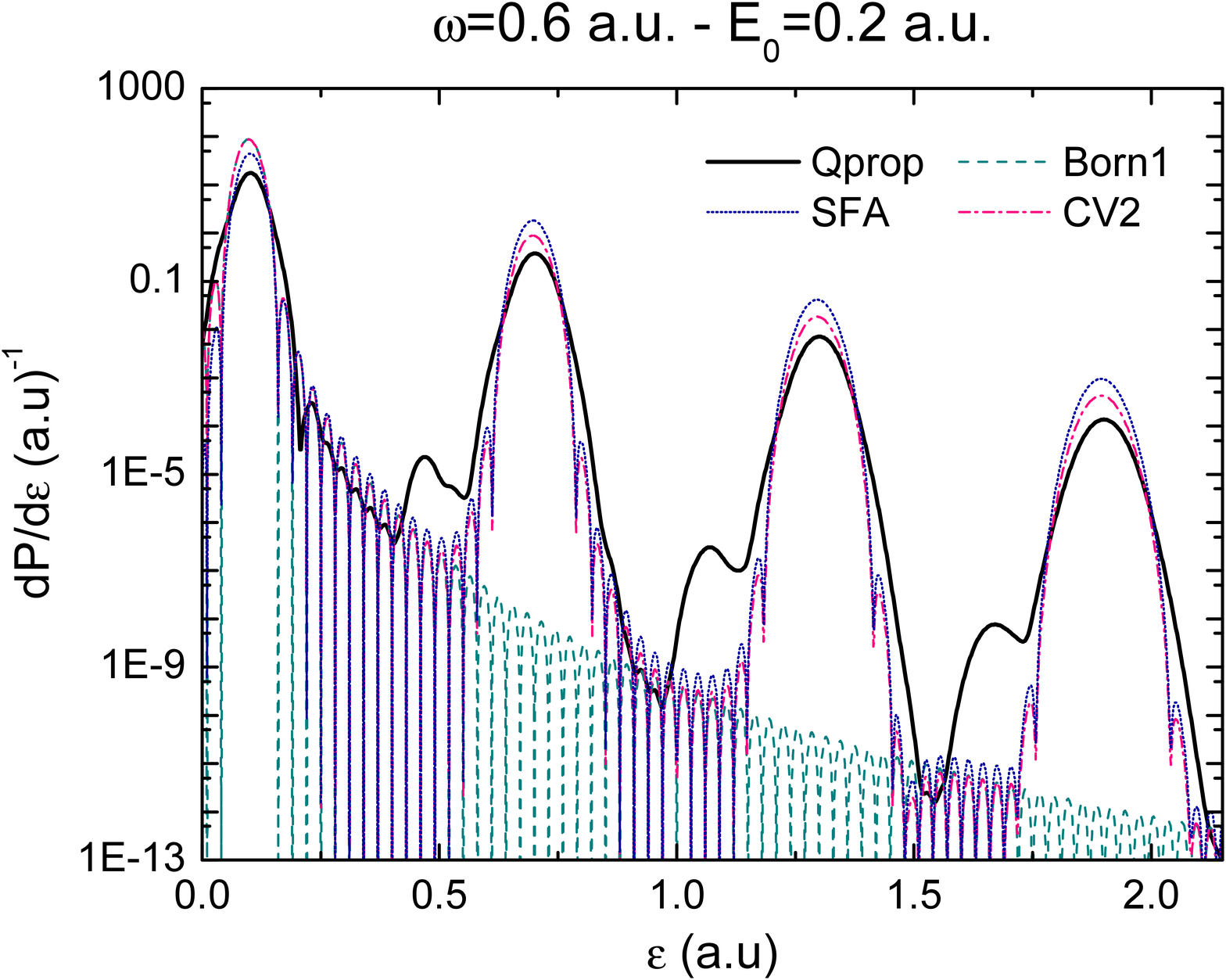}
\begin{caption}{Spectra computed under non-perturbative conditions: Laser pulse as in figure 2 but with field amplitude equal to $E_{0}=0.2$ a.u. ($1.41\times 10^{15} W/cm^{2})$.}
\end{caption}
\label{NonPertB1SFACV}
\end{figure}

In figure 3, spectra calculations with the same theories for non perturbative conditions are presented. In this case, the laser pulse has the same frequency and duration as the pulse of figure 2, but the field amplitude has been increased to $E_{0}=0.2$ a.u. ($1.41\times 10^{15}$ W/cm$^{2}$) to model a non perturbative condition.
From the comparison with exact TDSE results \cite{bauer06}, it can be observed that for this non-perturbative ionization regime, all the theories overestimate  not only the principal but also the secondary ATI peaks, as well as the background of the spectrum. As the field intensity increases, even though the computed peaks of the spectrum keep their appropriate positions, all the three theories overestimate the magnitude of the ATI peaks and the background of the spectrum.

\begin{figure}
\includegraphics[width=14 cm,height=9 cm]{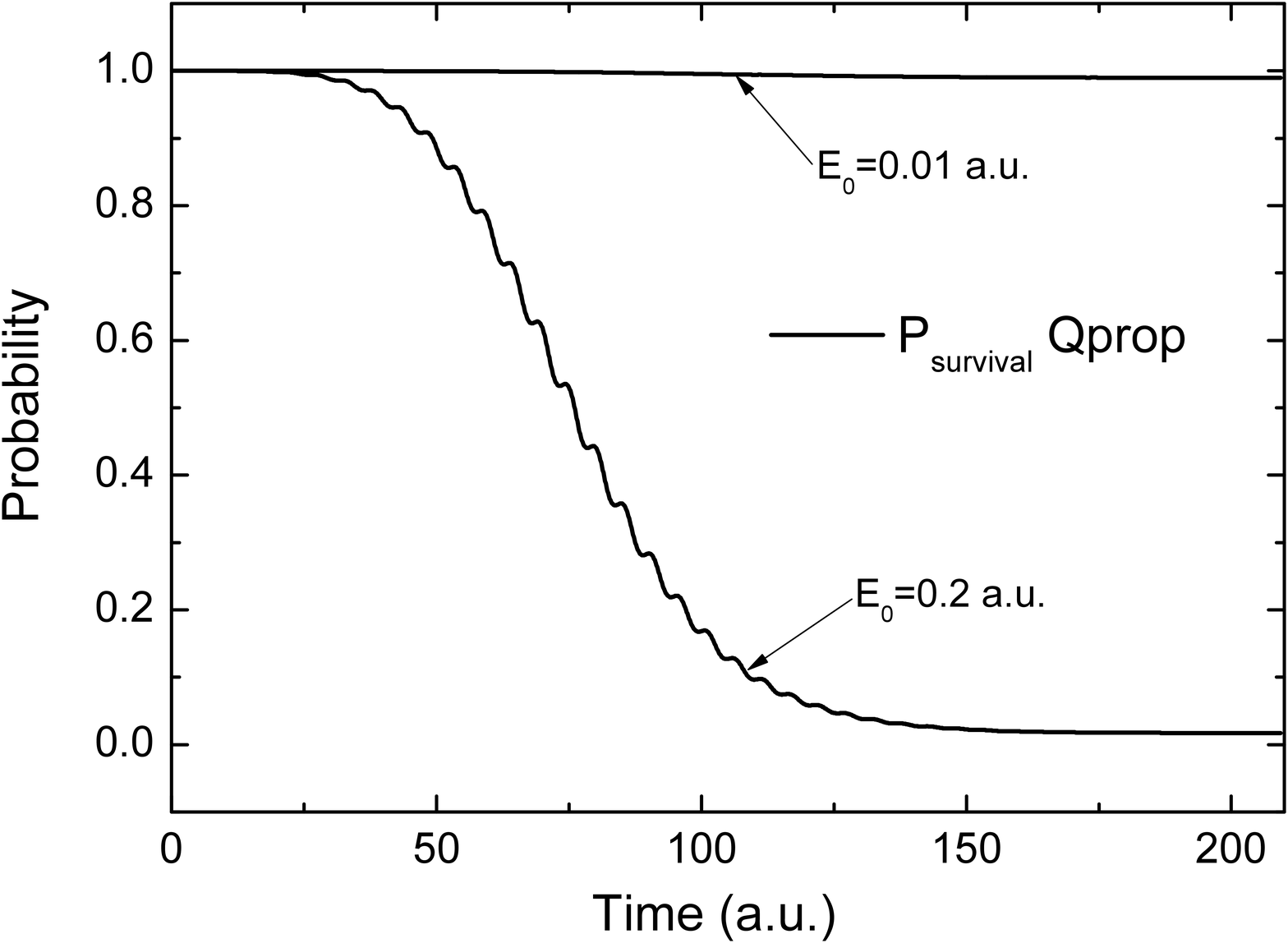}
\begin{caption}{Exact survival probabilities obtained with the Qprop code under perturbative conditions ($E_{0}=0.01$ a.u.) and non-perturbative conditions ($E_{0}=0.2$ a.u.). The laser frequency and number of cycles is the same as in figures 2 and 3.}
\end{caption}
\label{PertNonPertB1SFACV}
\end{figure}

The explanation behind this breakup of the theories considered in figure  3, lies in the quick depletion of the initial states when the conditions become non-perturbative. This feature in the ionization process is a consequence of the rise of the laser field amplitude as may be observed in figure 4, where the exact numerical survival probability computed with the Qprop code is plotted for both electric fields amplitudes.: $E_{0}=0.01$ a.u. (perturbative conditions) and $E_{0}=0.2$ a.u. (non-perturbative conditions). This figure shows that under perturbative conditions ($E_{0}=0.01$ a.u.), the depletion of the initial state is practically null, while in for the non-perturbative case ($E_{0}=0.2$ a.u.), the depletion of the initial state is  almost complete,  from the second half of the pulse duration.\\

\begin{figure}
\includegraphics[width = 14 cm,height = 9 cm]{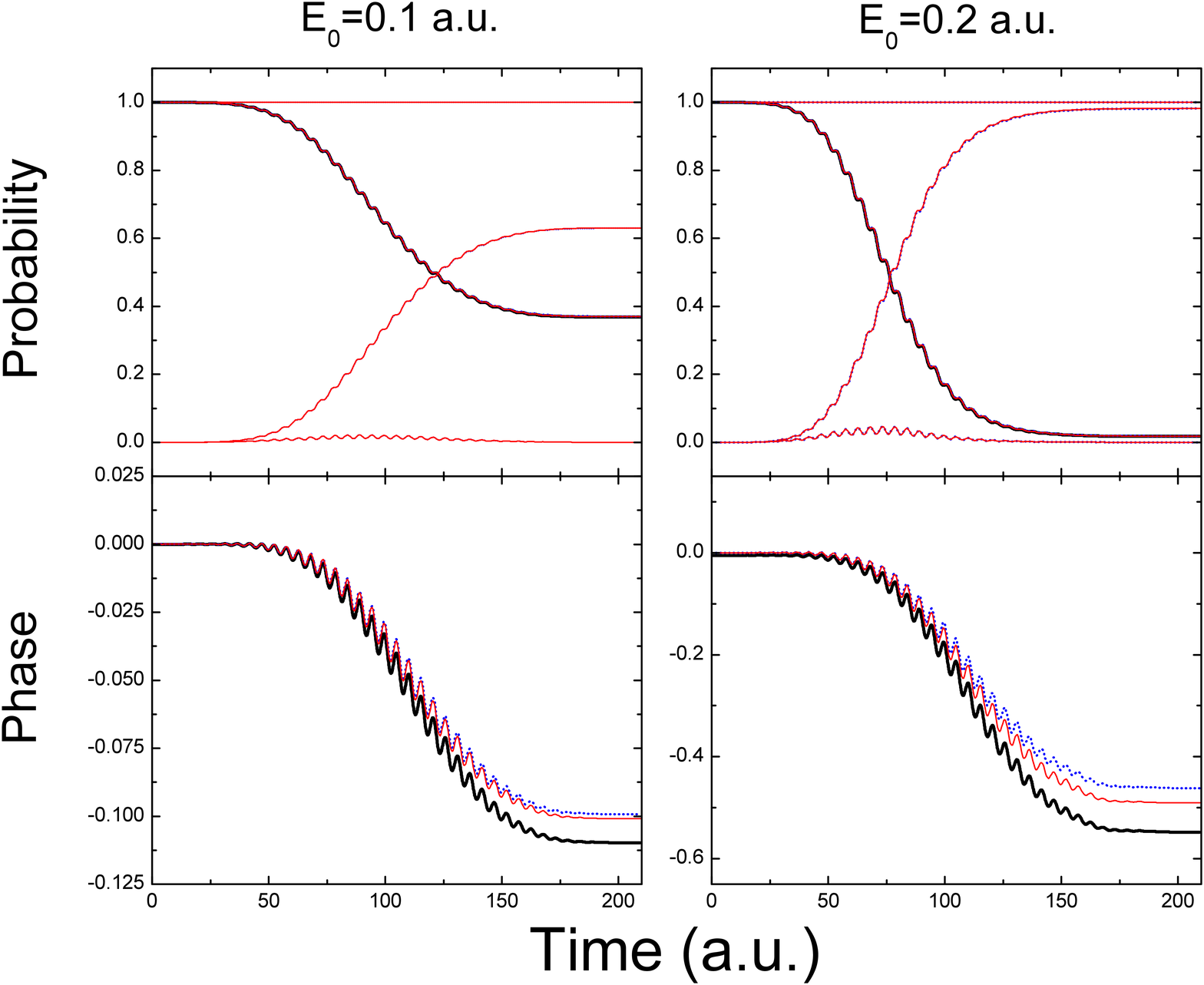}
\begin{caption}{Results obtained  for a laser pulse of frequency $\omega=0.6$ a.u.($16.33$ eV) and $20$ cycles of duration. Two laser field amplitudes are considered: $E_{0}=0.1$ a.u. (first column) and $E_{0}=0.2$ a.u. (second column).  In the first row, survival, total excitation, ionization, and total probabilities as a function of time are plotted: Results obtained with the Qprop code \cite{bauer06} (full wide line), and with the models $UM^{DC}$ (dotted line) and $UM^{DC+CC}$ (full thin line). In the second row of the figure, the corresponding amplitude phases are plotted.}
\end{caption}
\label{Probabilities1}
\end{figure}

More detailed information can be obtained from  the analysis of both, the module square (survival probability) and the phase of the initial state amplitude.

In figure 5, the results obtained for the two theories, $UM^{DC}$ (dotted lines) and   $UM^{DC+CC}$ (solid thin lines). Two field amplitudes, $E_{0}=0.1$ (first column) a.u. and $E_{0}=0.2$ a.u. (second column) are considered. The solid wide line gives the exact magnitudes computed with the Qprop code \cite{bauer06}. The survival probabilities computed with both theories, $UM^{DC}$  and $UM^{DC+CC}$, are in perfect agreement with the TDSE solution during all the pulse duration as can be appreciated in the figures. All three plots can not be distinguished among them. The depletion of the initial state is almost complete in the case with $E_{0}=0.2$ a.u. and only half of the initial state population depletion occurs in the case with $E_{0}=0.1$ a.u..

The remaining curves plotted in the first row of the figure are the total probability excitation to bound states (lower curves), the total ionization probability (rising curves), and the sum of all transition probabilities computed with the two models here introduced (constant curves). As shown in appendix A, our procedure is unitary and therefore the  sum of all the transition probabilities is one for all times as verified in the figure.
For these curves we present only the $UM^{DC}$  and   $UM^{DC+CC}$, which cannot be distinguished from each other.

In the second row of figure 5, the corresponding initial state amplitude phase for the two considered field intensities are shown. From the figures, it may be observed that the two models provide good account for the the exact numerical phases computed with the Qprop code, up to almost the second half of the laser pulse. During the last part of the pulse, departure from the exact phase initial transition amplitude is verified.

This is a consequence of the initial state depletion. Apart from the small quantitative difference between the two models here presented, all the time dependent quantities plotted in figure 5 behave in a similar way. Only quantitative differences are observed.

\begin{figure}
\includegraphics[width=14 cm,height=9 cm]{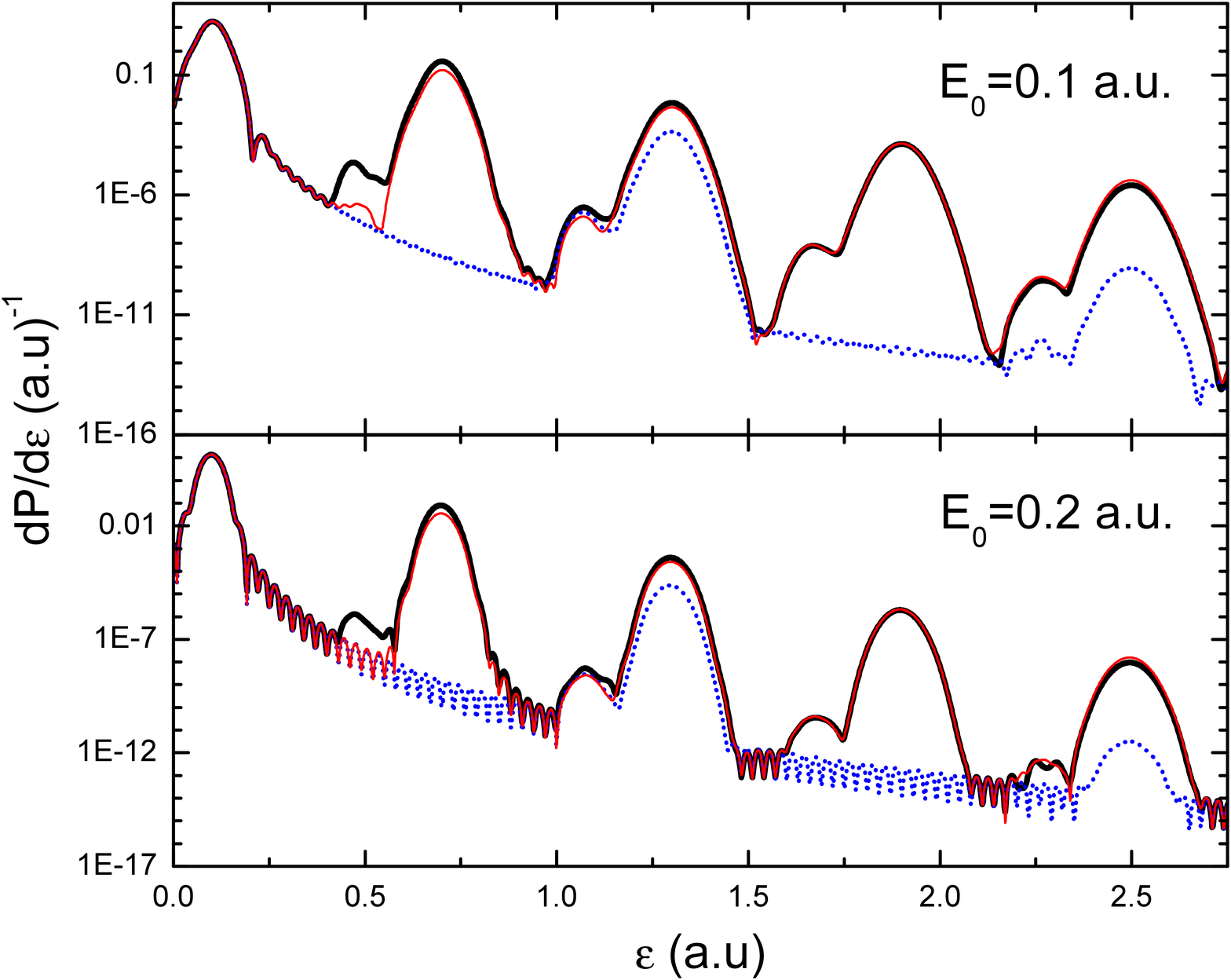}
\begin{caption}{ATI spectra: exact TDSE obtained with the Qprop code \cite{bauer06} (full wide line), $UM^{DC}$ (dotted line) and $UM^{DC+CC}$ (full thin line). The odd peaks are only featured by the theory accounting for continuum-continuum coupling.}
\end{caption}
\label{Spectra1}
\end{figure}

 The importance of including or not the coupling between continuum states becomes apparent only when the energy spectra computed with the two theories are plotted (see figure 6) . The absence of coupling between continuum states in the simplified version, prevents the emergence of the even peaks of the spectrum. In this case, the spectrum does not show the even ATI peaks as these peaks are populated with $s$ and $d$ electrons, which are  not accounted for in the $UM^{DC}$ model since only the $p$ states are coupled with the initial ground state.

In figure 6, with the same laser pulse conditions of figure 5, the spectra corresponding to theories $UM^{DC}$ (dotted lines) and $UM^{DC+CC}$ (solid thin lines) are plotted. The spectrum computed with the Qprop code is also plotted in the figure (full wide line).
Both models display the first ionization peak in agreement with the corresponding exact TDSE solution�s one, but in the case of the model without the continuum-continuum coupling, only the odd ATI peaks are displayed, as expected. Further, in this case,  the magnitude of the shown peaks is underestimated although their positions are the correct ones. The most complete  model gives a good approximation to the magnitude of all the peaks of the spectra. It also accounts for the intermediate structures  appearing near each of the higher order peaks of the spectra, with the exception of the very first structure which is traced back to the other discrete-discrete coupling not accounted by the $UM^{DC+CC}$ model ($p-s$ and $p-d$ transitions). The background corresponding to the exact calculations is well reproduced by both models.

\begin{figure}
\includegraphics[width=14 cm,height=9 cm]{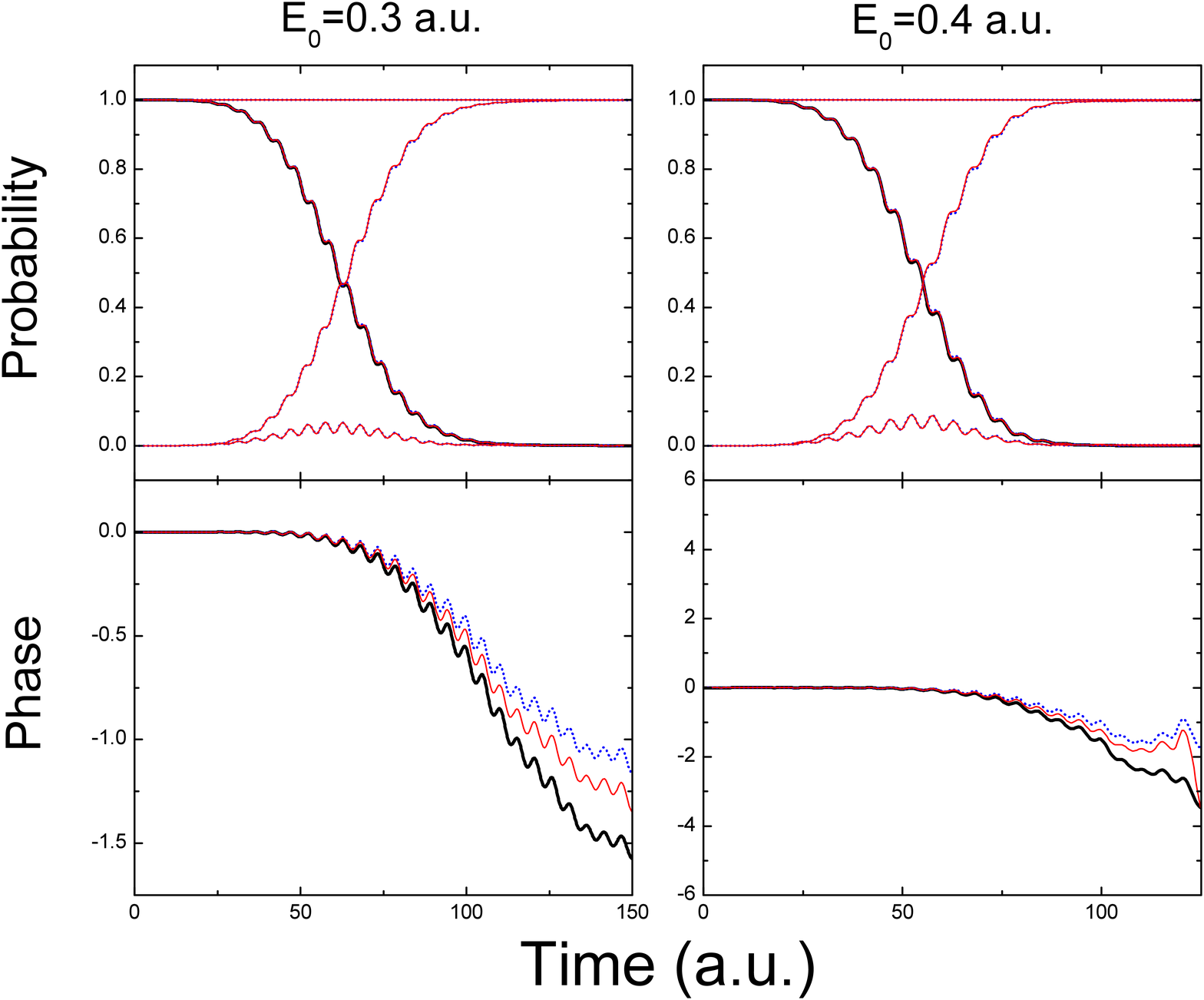}
\begin{caption}{Idem figure 5 but with two larger field amplitudes: $E_{0}=0.3$ a.u. (first column) and $E_{0}=0.4$ a.u.(second column).}
\end{caption}
\label{Probabilities2}
\end{figure}

In figure 7, the same quantities as in figure 5 are plotted but this time for greater laser field intensities, keeping the same laser pulse frequency and duration. The results for $E_{0}=0.3$ a.u. and for $E_{0}=0.4$ a.u. are presented in column one and two respectively. In this case, the same overall behavior is observed for the probability amplitudes and the initial state amplitude phases as a function of time. As expected, depletion of the initial state is achieved earlier than in previous cases with  lower intensities (figure 5).

For the higher laser field intensities in figure 7, only a fraction of the laser pulse duration is displayed as full depletion has already been achieved. Once the pulse is fully depleted phase amplitudes become meaningless.

\begin{figure}
\includegraphics[width=14 cm,height=9 cm]{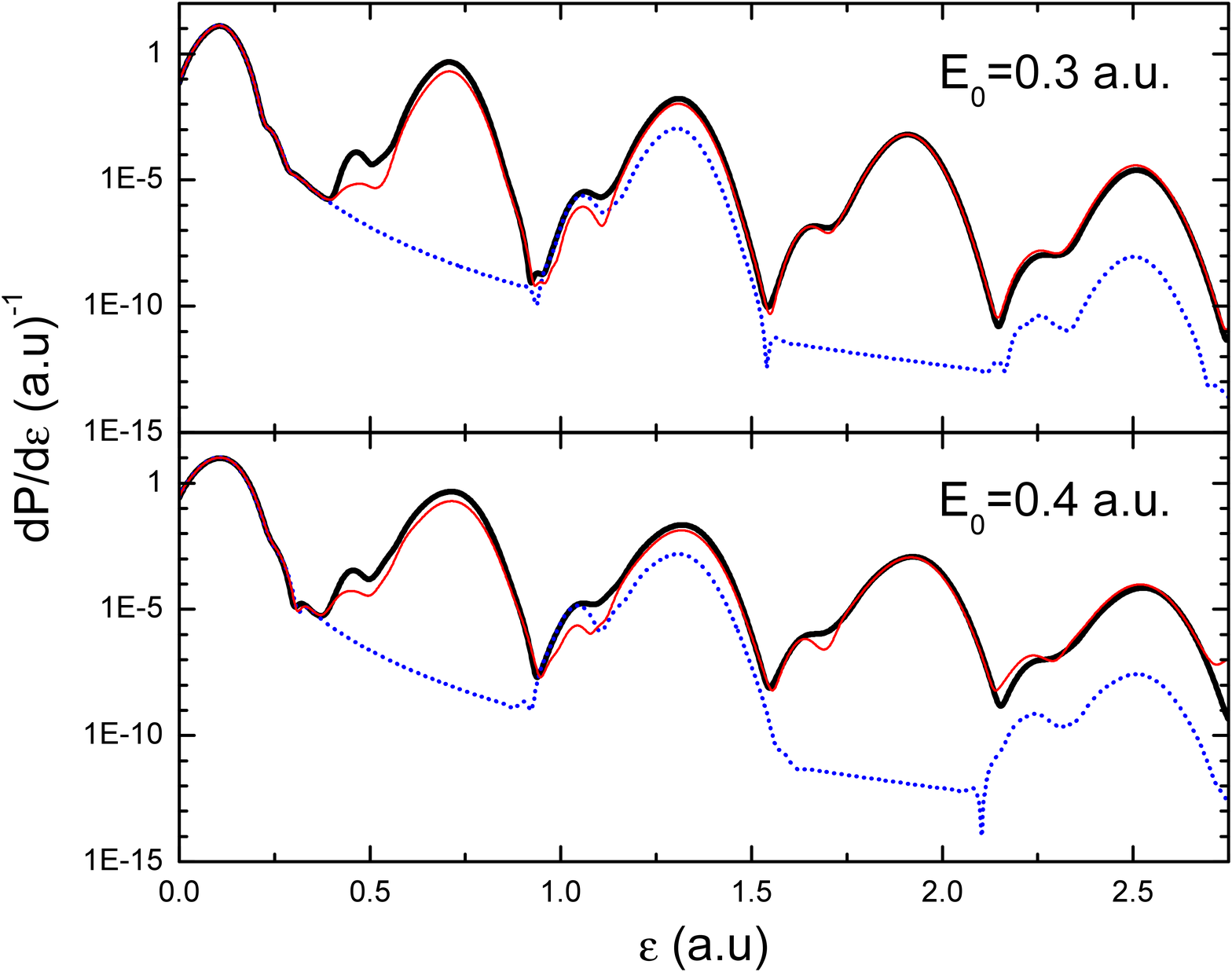}
\begin{caption}{ATI spectrum: exact TDSE obtained with the Qprop code \cite{bauer06} (full wide line), $UM^{DC}$ (dotted line) and $UM^{DC+CC}$ (full thin line). The odd peaks are only featured by the full coupled theory.}
\end{caption}
\label{Spectra2}
\end{figure}

The corresponding spectra for the laser parameters of figure 7 are presented in figure 8. As in the case of figure 6, with smaller laser field amplitudes, the spectra computed with the simplified version of the model only display the odd peaks as discussed. Again, the magnitude of the peaks of higher order than the first is shown to have been underestimated by the simplified version, even though, its position appears to be correct.

\section{Conclusions}

Both versions of the  unitary theory presented in this work account well for the depletion of the initial state under non perturbative conditions.
Even though both theories give the correct description of the first ATI peak of the spectrum, only the full coupled $UM^{DC+CC}$ provides the correct description of all the ATI peaks of the spectrum, in particular of the odd peaks that the simplified version can not  reproduce.\\
The theory shares the unitary property with the exact solution.
The theory is much simpler than the exact numerical solution of the TDSE: A simple integro-differential equation for the initial state amplitude has to be solved, and the remaining probability amplitudes are obtained straightforwardly from the solution of the IDE.

\begin{acknowledgments}
This work was partially supported by the  Consejo Nacional de Investigaciones Cient\'{\i}ficas y T\'{e}cnicas PIP  and Universidad de Buenos Aires UBACYT, Argentina.
\end{acknowledgments}

\appendix*

\section{}

For the sake of simplicity we define two functions $A_{i}$ and $\gamma_{Ci}^{V}$ by	
\begin{eqnarray}
\qquad \qquad \qquad A_{i}(t)=A(t)\exp(i\varepsilon_{i})\\
\qquad \gamma_{Ci}^{V}(t)=-\int_{0}^{t}dt^{\prime }A(t^{\prime})\exp[-i\varepsilon_{i}t^{\prime}]a_{i}(t^{\prime})h_{C}^{V}(t-t^{\prime}),
\label{AuxFunc}
\end{eqnarray}
\\
in these terms the ionization rate Eq. (\ref{DPioni1}) takes the form
\begin{equation}
\qquad \qquad \qquad \qquad \qquad \dot{P}_{ioni}(t) = - A_{i}(t) a_{i}^*(t) \gamma_{Ci}^{V}(t) + c.c.
\label{DPioni2}
\end{equation}
or in other words,
\begin{equation}
\qquad \qquad \qquad \qquad \qquad \dot{P}_{ioni}(t)= - 2 \texttt{Re}[(A_{i}(t) a_{i}^*(t) \gamma_{Ci}^{V}(t))].
\label{DPioni3}
\end{equation}
The corresponding probability of the system to be found in the initial state is given by
\begin{equation}
\qquad \qquad \qquad \qquad \qquad \qquad P_{i}(t)=a_{i}(t)a^*_{i}(t),
\label{Pi}
\end{equation}
and then
\begin{equation}
\qquad \qquad \qquad \qquad \qquad \dot{P}_{i}(t)=\dot{a}_{i}(t)a^*_{i}(t)+ c.c =2 \texttt{Re}[\dot{a}_{i}(t)a^*_{i}(t)]
\label{DPi}
\end{equation}
The amplitude of the initial state may be written as
\begin{equation}
\qquad \qquad \qquad \qquad \qquad \qquad a_{i}(t)=A_{i}(t) \gamma_{i}^{V}(t)
\label{ai}
\end{equation}
where $\gamma_{i}^{V}(t)$ is defined in analogy with $\gamma_{Ci}^{V}(t)$ in terms of $h(t-t^{\prime})$ instead of $h_{C}(t-t^{\prime})$ and then
\begin{equation}
\qquad \qquad \qquad \qquad \qquad \qquad \dot{P}_{i}(t)=2 \texttt{Re}[A_{i}(t)a_{i}^*(t)\gamma_{i}^{V}(t)]	
\label{DPi1}
\end{equation}
Furthermore,
\begin{equation}
\qquad \qquad \qquad \qquad \qquad \qquad P_{n}(t)=a_{n}(t)a^*_{n}(t),
\label{Pn}
\end{equation}
and then
\begin{equation}
\qquad \qquad \qquad \qquad \qquad \qquad \dot{P}_{n}(t)=2 \texttt{Re}[\dot{a}_{n}(t)a^*_{n}(t)]
\label{DPn}
\end{equation}
where from $\dot{a}_{n}(t)$ given in Eq. (\ref{DnAmplitude}),
\begin{equation}
\qquad \qquad \qquad \qquad \qquad \dot{a}^*_{n}(t)= i A(t)p_{z_{ni}} \exp[i(\varepsilon_{i}-\varepsilon_{n})t]a^*_{i}(t).
\label{Danconj}
\end{equation}
Then, from Eq. (\ref {nAmplitude1}) and the last equation
\begin{equation}
a_{n}(t)\dot{a}^*_{n}(t)=|p_{z_{ni}}|^2 A(t) \exp[i(\varepsilon_{i}-\varepsilon_{n})t]a^*_{i}(t)\int_0^{t}dt^{\prime} A(t^{\prime}) \exp[-i(\varepsilon_{i}-\varepsilon_{n})t^{\prime}]a_{i}(t^{\prime})
\label{anDanconj}
\end{equation}
And in consequence,
\begin{equation}
\sum_{n\neq i} a_{n}(t)\dot{a}^*_{n}(t)=\sum_{n\neq i}|p_{z_{ni}}|^2 A(t) \exp[i(\varepsilon_{i}-\varepsilon_{n})t]a^*_{i}(t)\int_0^{t}dt^{\prime} A(t^{\prime}) \exp[-i(\varepsilon_{i}-\varepsilon_{n})t^{\prime}]a_{i}(t^{\prime})
\label{SumanDanconj1}
\end{equation}
Which means that
\begin{equation}
\qquad \qquad \qquad \qquad \qquad \sum_{n\neq i} a_{n}(t)\dot{a}^*_{n}(t)= - A_{i}(t)a_{i}^*(t)\gamma_{Di}^{V}(t)
\label{SumanDanconj2}
\end{equation}
or what is equivalent
\begin{equation}
\qquad \qquad \qquad \qquad \qquad \sum_{n\neq i}\dot{P}_n(t) =-2 \texttt{Re}[A_{i}(t)a_{i}^*(t)\gamma_{Di}^{V}(t)].
\label{DPD}
\end{equation}
Then from the last equation together with Eq. (\ref{DPi1}) and Eq. (\ref{DPioni3}) it is clear that
\begin{equation}
\qquad \qquad \qquad \qquad \dot{P}_{i}(t) + \sum_{n \neq i}\dot{P}_n(t) + \dot{P}_{ioni}(t) =\dot{P}_{TOTAL}(t)= 0.
\label{DPimasSumDPnmasDPioni}
\end{equation}
This means that $P_{TOTAL}(t)$	is constant in the interval $(0,\tau)$ and because $P_{TOTAL}(0)=1$, then $P_{TOTAL}(t)=1$ $\forall t \in (0,\tau)$.
In other words, the model is \emph{unitary}.



\end{document}